\documentclass[11pt,a4paper]{article}
\usepackage{jcappub}
\usepackage{amsmath,amssymb,enumerate,epsfig,subfigure,caption2
}
\usepackage{overpic}
\usepackage{leftidx}
\usepackage{subfigure}
\usepackage[english]{babel}
\usepackage{amsmath}
\usepackage{amssymb}
\usepackage{epsfig}
\usepackage{graphics,psfrag,rotating}
\usepackage{graphicx}
\usepackage{dcolumn}
\usepackage{bm}%

\def\3nab{\tilde{\nabla}}

\def\be {\begin{equation}}
\def\ee {\end{equation}}
\def\bea {\begin{eqnarray}}
\def\eea {\end{eqnarray}}

\def\3nab{\tilde{\nabla}}

\def\hsp5{\hspace{5mm}}

\def\case#1/#2{\textstyle\frac{#1}{#2}}

\def\be{\begin{equation}}
\def\ee{\end{equation}}
\def\bea{\begin{eqnarray}}
\def\eea{\end{eqnarray}}

\def\nn{\nonumber \\}

\def\case#1/#2{\textstyle\frac{#1}{#2} }

\title{Junction conditions in extended Teleparallel gravities}

\author{
\'Alvaro de la Cruz-Dombriz$^{a}\, \footnote{ dombriz [at] fis.ucm.es}$ 
Peter K. S. Dunsby$^{b,c}\,\footnote{ peter.dunsby [at] uct.ac.za}$, 
Diego S\'aez-G\'omez$^{b,d}\,\footnote{ diego.saezgomez [at] uct.ac.za}$
}
\affiliation{
$^{a}$Departamento de F\'{\i}sica
Te\'orica I, Ciudad Universitaria, Universidad Complutense de Madrid, E-28040 Madrid,
Spain.\\
$^{b}$  Astrophysics, Cosmology and Gravity Centre (ACGC), Department of Mathematics and Applied Mathematics, University of Cape Town, Rondebosch 7701, Cape Town, South Africa.\\
$^{c}$ South African Astronomical Observatory,  Observatory 7925, Cape Town, South Africa\\
$^{d}$ Fisika Teorikoaren eta Zientziaren Historia Saila, Zientzia eta Teknologia Fakultatea,\\
Euskal Herriko Unibertsitatea, 644 Posta Kutxatila, 48080 Bilbao, Spain.\\
}

\date{\today}

\abstract{
In the context of extended Teleparallel gravity theories, we address the issue of junction conditions required to
guarantee the correct matching of different regions of spacetime. In the absence of shells/branes, these conditions turn out to be more restrictive than their counterparts in General Relativity as in other extended theories of gravity. In fact, the general junction conditions on the matching hypersurfaces depend on the underlying theory and a new condition
on the induced tetrads in order to avoid delta-like distributions in the field equations.
This result imposes strict consequences on the viability of standard solutions such as the Einstein-Straus-like construction. We find that the continuity of the scalar torsion is required  in order to recover the usual General Relativity results.}

\begin{document}
\maketitle

\section{Introduction}

Teleparallel gravity is a gravitational theory for the translation group, associating a Minkowskian tangent space to every point of the spacetime. Teleparallel gravity 
parallelly transports the so-called vierbeins/tetrads field, providing the name of the theory ({\it c.f.} \cite{Teleparallelism} for a comprehensive review). In order to achieve this,  the theory is constructed in terms of the so-called Weitzenb\"ock connection instead of the Levi-Civita connection. Unlike the Levi-Civita connection, the Weitzenb\"ock connection is not commutative under the exchange of the lower indices, which not only induces a non-zero torsion, but also a zero Ricci scalar.  
%
%
%
%
Thus, the covariant action for Teleparalell gravity is constructed in terms of the scalar torsion $T$. 
A relevant feature of this theory lies in the fact that  every solution of General Relativity (GR) is also a solution for Teleparallel gravity. 
Moreover, some extensions of Teleparallel gravity in the cosmological context have attracted some interest over the last few years. This is due to the fact that Dark Energy may be described within the framework of these theories ({\it c.f. } \cite{Bengochea:2008gz}).

%
In analogy of higher-order theories of gravity, such as $f(R)$ theories \cite{fR_varia}, Teleparallel gravity has been extended by constructing gravitational Lagrangians in terms of more complex functions of the torsion scalar. They have been referred to as $f(T)$ theories. When introducing extra terms in the action, the aforementioned equivalence between GR and Teleparallel gravity now does not remain between $f(T)$ and $f(R)$ theories. 
However,  the $f(T)$ gravitational field equations are second order whereas modifications of General Relativity, such as $f(R)$ theories, are usually higher order. This results for instance in interesting differences, because no extra gravitational waves modes (in comparison to GR) appear  \cite{Bamba:2013ooa}. Furthermore, the inflationary scenario has been studied in these theories \cite{Ferraro:2006jd}, as well as the non-invariance of the theory under local Lorentz transformations \cite{Li:2010cg}. In addition, some other extensions of Teleparallel gravity have also been proposed,  such as conformally invariant actions in Teleparallel gravity \cite{Bamba:2013jqa}. 

%
As it is widely known, extensions of Teleparallel gravity are not invariant under local Lorentz transformations (see Ref.~\cite{Li:2010cg}). Thus, the field equations will be sensitive to the choice of tetrads and consequently,  the determination of the correct tetrad fields, leading to a metric tensor with some desirable symmetries, has attracted some attention in the last few years. The key-point lies in the requirement of a correct parallelisation of spacetime, i.e., the correct choice of tetrads, in order to guarantee a solution of the field equations while recovering the desired metric.
In fact,  the choice of some tetrads may impose restrictions in the form of the 
$f(T)$ theories able to satisfy the field equations. Authors in \cite{good-and-bad} have claimed that once good tetrads have been chosen the functional form of $f(T)$ must not be restricted. Literature devoted to this subtlety in Teleparallel gravity has mainly focused on standard solutions such as Friedmann-Lema\^itre-Robertson-Walker (FLRW) universes \cite{ref27} or  vacuum static and spherically symmetric (Schwarzschild) solution \cite{Argentinos}. Considerable attention has been devoted to the Schwarzschild solution because of its importance \cite{SSS_references,29,33,34, Boehmer:2011gw}. The use of naive tetrad fields which were expected to represent a Schwarzschild solution were shown to be inconsistent with this solution \cite{Argentinos}, \cite{29}-\cite{34}. Nevertheless,  a set of tetrads (the so-called {\it Schwarzschild frame}), able to cover this solution can be obtained by performing a convenient boost (or rotation \cite{good-and-bad}) of the isotropic set of tetrads 
\cite{Argentinos}.  Moreover, in the context of $f(T)$ theories the existence of vacuum solution with both vanishing $R$ and $T$ leads to $f(T)=T+ \mathcal{O}(T^2)$, for which the Schwarzschild tetrads lead to vacuum solutions. In addition Birkhoff's theorem has been addressed after several attempts \cite{good-and-bad, 32, Boehmer:2011gw} leading to Schwarzschild-de Sitter solution with applications in 3-dimensional scenarios \cite{Argentinos}.

Cosmological solutions in Teleparallel theories have also been given extensive coverage in the literature  \cite{Bengochea:2008gz}.  Analogously to the subtleties about the correct choice of tetrads in static scenarios,  the use of diagonal tetrad in spherical coordinates in $f(T)$ theories, 
constrains the gravitational Lagrangian, since $f_{TT}=0$ is a necessary condition to fulfill the field equations. In other words, the $f(T)$ Teleparallel gravity would reduce to GR. 
Nevertheless, rotation of these tetrads shows that no constraints on the $f(T)$ action appear and consequently, general classes of $f(T)$ models are allowed, in principle in principle allowed \cite{good-and-bad}.
The fact that rotations suffice to find adequate tetrads providing either spherically symmetric and static solutions or homogeneous and isotropic solutions, opens up the possibility of considering more general local Lorentz transformations guaranteeing the existence of well-behaved tetrads.

The application of junction conditions is a key issue in every gravitational theory aiming to describe satisfactorily, among other things, stellar configurations or collapsing bodies. Since these conditions guarantee a smooth transition through the matching hypersurface between different spacetime regions, their violation poses serious shortcomings in any theory under consideration \cite{Israel,Mars:1993mj,Mars:2013ooa}. Previous work dealing with other types of extended theories of gravity demonstrated that additional junction conditions with respect to standard GR conditions are required. A case in point is $f(R)$ gravity, where the scalar (Ricci) curvature has to be continuous across the matching hypersurface when only jumps in the matter sector are permitted ({\it c.f.}  \cite{Deruelle:2007pt,Clifton:2012ry,Senovilla:2013vra}). This result implies that matched solutions in Einstein gravity will not necessarily exist in the context of fourth-order $f(R)$ gravity due to the presence of these additional junction conditions. 
With respect to Teleparallel gravity, i.e., $f(T)=T$, this theory does recover the standard GR results, and consequently this fact demonstrates that the usual Israel junction conditions \cite{Israel} are satisfied in this case independently of the tetrads, as the theory remains invariant under local Lorentz transformations. 
Nevertheless, this is not the case in $f(T)$ theories with  $f(T)\neq T$, where the matching 
 between two different spacetime regions provides junction conditions different from those in GR.
 In this paper we shall demonstrate 
that the presence of 
a new junction condition, which is absent in Teleparallel gravity, plays a crucial role in the possibility of matching two different spacetime regions. 
In the absence of branes, i.e., just considering the case of two different regions of spacetime described by different metrics, the new conditions are $f(T)$ model-dependent and require the continuity of the induced tetrad. It is not straightforward to combine both requirements in order to recover usual classical solutions. However,  to illustrate this difficulty, we attempt to construct the Einstein-Straus model, where an empty spherically symmetric region is matched to a general FLRW metric \cite{Einstein-Straus}. Unlike the case of $f(R)$ gravity where the condition on the Ricci scalar does not allow one to reconstruct the Einstein-Straus model ({\it c.f.} \cite{Clifton:2012ry}), we shall show how this model may be reconstructed in $f(T)$ gravity by an adequate procedure. 

The paper is organised as follows: in Section \ref{sec2} a brief introduction to $f(T)$ theories is given. Section \ref{sec3} deals with the Arnowitt-Deser-Misner (ADM) decomposition in these theories. This decomposition, when a trivial redefinition of coordinates is performed, will enable us to provide the key equations of our analysis in a transparent way once the spacetime foliation is performed in the correct coordinates. Then, the junction conditions for $f(T)$ gravity are derived in Section \ref{sec4}. Section \ref{sec5} is devoted to the reconstruction of the Einstein-Straus model within these theories. We conclude in Section \ref{Sec6} with a discussion about possible applications of our results.

Throughout the paper we shall follow the following conventions: $D_{\mu}$ will represent the covariant derivative with respect to the Levi-Civita connection $\Gamma^{\alpha}_{\mu\nu}$ and $\stackrel{\bullet}{D}_{\mu}$ holds for the covariant derivative in terms of the Weitzenb\"ock connection $\stackrel{\bullet}{\Gamma^{\alpha}}_{\mu\nu}$. 
Greek indexes such as $\mu, \nu...$ refer to spacetime indexes, the latin letters $a, b, c...$ refer to the tetrads indexes associated with the tangent space, while the latin letters $i, j, k...$ express indexes of a 3-hypersurface of the spacetime.

\section{$f(T)$ gravity} \label{sec2}

Teleparallel gravity can be described as an orthonormal basis of tetrads  $e_{a}(x^{\mu})$  defined in the tangent space at every point $x^{\mu}$ on the spacetime. Then, the metric can be expressed in terms of the tetrads as follows
\begin{eqnarray}
{\rm d}s^{2} &=&g_{\mu\nu}{\rm d}x^{\mu}{\rm d}x^{\nu}=\eta_{ab}\theta^{a}\theta^{b}\label{1}\; ,\\
{\rm d}x^{\mu}& =&e_{a}^{\;\;\mu}\theta^{a}\; , 
\quad 
\theta^{a}=e^{a}_{\;\;\mu}{\rm d}x^{\mu}\label{2}\; ,
\label{1.1}
\end{eqnarray} 
where $\eta_{ij}=\text{diag}(+1,-1,-1,-1)$ and $e_{a}^{\;\;\mu}e^{a}_{\;\;\nu}=\delta^{\mu}_{\nu}$ or $e_{a}^{\;\;\mu}e^{b}_{\;\;\mu}=\delta^{b}_{a}$.  The tetrads $e^{a}_{\;\;\mu}$ represent the dynamic fields of the theory.
\par
Furthermore, in Teleparallel gravities, the connection describing the covariant derivatives of tensors
is given by the Weitzenb\"{o}ck connection instead of the Levi-Civita connection:
\begin{eqnarray}
\stackrel{\bullet}{\Gamma^{\alpha}}_{\mu\nu}=e_{a}^{\;\;\alpha}\partial_{\nu}e^{a}_{\;\;\mu}=-e^{a}_{\;\;\mu}\partial_{\nu}e_{a}^{\;\;\alpha}\label{co}\; ,
\label{WC}
\end{eqnarray}
which leads to a vanishing scalar curvature but non-zero torsion. Then, the torsion tensor $T^{\alpha}_{\;\;\mu\nu}$ is defined as the antisymmetric part of the connection (\ref{WC}):
\begin{eqnarray}
T^{\alpha}_{\;\;\mu\nu}&=&\stackrel{\bullet}{\Gamma^{\alpha}}_{\nu\mu}-\stackrel{\bullet}{\Gamma^{\alpha}}_{\mu\nu}=e_{a}^{\;\;\alpha}\left(\partial_{\mu} e^{a}_{\;\;\nu}-\partial_{\nu} e^{a}_{\;\;\mu}\right)\ , 
\label{tor}
\end{eqnarray}
while the contortion 
tensor $K^{\mu\nu}_{\;\;\;\;\alpha}$ is defined as the difference between the Weitzenb\"{o}ck and the Levi-Civita connections:
\be
K^{\alpha}_{\;\; \mu\nu}= \stackrel{\bullet}{\Gamma^{\alpha}}_{\mu\nu}-\Gamma^{\alpha}_{\mu\nu}=\frac{1}{2}\left(T_{\mu\;\;\ \nu}^{\;\; \alpha}+T_{\nu\;\;\ \mu}^{\;\; \alpha}-T_{\;\; \mu \nu}^{\alpha}\right)\;,
\label{contor2}
\ee
or equivalently,
\begin{eqnarray}
K^{\mu\nu}_{\;\;\;\;\alpha}&=&-\frac{1}{2}\left(T^{\mu\nu}_{\;\;\;\;\alpha}-T^{\nu\mu}_{\;\;\;\;\alpha}-T_{\alpha}^{\;\;\mu\nu}\right)\ .
\label{contor}
\end{eqnarray}
Furthermore, in order to construct the gravitational action, a tensor $S_{\alpha}^{\;\;\mu\nu}$ is defined, by using the torsion and contortion tensors:
\begin{eqnarray}
S_{\alpha}^{\;\;\mu\nu}&=&\frac{1}{2}\left( K_{\;\;\;\;\alpha}^{\mu\nu}+\delta^{\mu}_{\alpha}T^{\beta\nu}_{\;\;\;\;\beta}-\delta^{\nu}_{\alpha}T^{\beta\mu}_{\;\;\;\;\beta}\right)\label{s}\;,
\end{eqnarray}
whereas the torsion scalar is constructed contracting the torsion tensor (\ref{tor}) and the tensor (\ref{s}) as follows,
\begin{eqnarray}
T=T^{\alpha}_{\;\;\mu\nu}S^{\;\;\mu\nu}_{\alpha}\label{te}\; ,
\label{scalar-torsion}
\end{eqnarray}
Alternatively, the torsion scalar can also be expressed in terms of the torsion tensor:
\be
T=\frac{1}{4}T^{\lambda}_{\;\;\;\mu\nu}T_{\lambda}^{\;\;\;\mu\nu}+\frac{1}{2}T^{\lambda}_{\;\;\;\mu\nu}T_{\;\;\;\;\;\lambda}^{\nu\mu}-T^{\rho}_{\;\;\;\mu\rho}T_{\;\;\;\;\;\nu}^{\nu\mu}\ .
\label{scalar-torsion2}
\ee
Then, the well-known action for Teleparallel gravity is given by~\cite{Teleparallelism} 
\begin{eqnarray}
\label{action}
S_G=\frac{1}{2\kappa^2}\int e\ T\ {\rm d}^4x\; .
\label{TeleAction}
\end{eqnarray}
In analogy of $f(R)$ gravity, this action can be further generalised into more general functions of scalar torsion:
\begin{eqnarray}
\label{action}
S_G=\frac{1}{2\kappa^2}\int e\ f(T){\rm d}^4x\; .
\label{ftaction}
\end{eqnarray}
By assuming a matter action $S_\mathrm{m}=\int e\ \mathcal{L}_\mathrm{m} {\rm d}^4x$ with 
$\mathcal{L}_\mathrm{m}$ being the matter Lagrangian, the field equations can be obtained  by varying the action (\ref{ftaction}) 
with respect to the tetrads, yielding 
\begin{eqnarray}
S^{\;\;\nu\rho}_{\mu}\partial_{\rho}Tf_{TT}+\left[e^{-1}e^{i}_{\mu}\partial_{\rho}\left(ee^{\;\;\alpha}_{i}S^{\;\;\nu\rho}_{\alpha}\right)+T^{\alpha}_{\;\;\lambda\mu}S^{\;\;\nu\lambda}_{\alpha}\right]f_{T}+\frac{1}{4}\delta^{\nu}_{\mu}f=\frac{\kappa^2}{2}\mathcal{T}^{\nu}_{\mu}\label{em}\; ,
\end{eqnarray}
where $\mathcal{T}^{\nu}_{\mu}=\frac{e_{a}^{\;\;\nu}}{e}\frac{\delta\mathcal{L}_\mathrm{m}}{\delta e_{a}^{\;\;\mu}}$ holds for the matter energy-momentum tensor, $f_{T}={\rm d} f(T)/{\rm d} T$ and $f_{TT}={\rm d}^{2} f(T)/{\rm d}T^{2}$.  In addition, by using the contortion tensor (\ref{contor2}), the relation between the Ricci scalar and the torsion scalar yields,
\be
R\,=\,-T-2D_{\mu}T^{\nu\mu}_{\;\;\;\;\;\nu}\ .
\label{RS}
\ee
Thus, it is straightforward to check that Teleparallel gravity is equivalent to GR, since the  covariant derivative in (\ref{RS}) can be removed in the action (\ref{TeleAction}). Furthermore, the relation (\ref{RS}) enables us to rewrite the field equations (\ref{em}) in the more usual covariant description,
\begin{equation} 
f_{T} \left( R_{\mu\nu} - \frac{1}{2} g_{\mu\nu} R \right) 
+ \frac{1}{2}g_{\mu\nu} \left( f(T) - f_{T} T \right)
+ 2f_{TT} S_{\nu\mu\rho} D^{\rho} T  = {\kappa}^2 
{\mathcal{T}}_{\mu\nu}\,, 
\label{covEq}
\end{equation}
where $R_{\mu\nu}$ and $R$ the Ricci tensor and Ricci scalar,  respectively. Let us stress at this stage that by setting $f(T) = T$, the field equations (\ref{covEq}) reduce to the standard Einstein field equations. 

\section{ADM decomposition in Teleparallel gravity}
\label{sec3}
In order to determine the junction conditions between different regions of a spacetime for the theory studied in this paper,  it is convenient  to 
foliate the spacetime in hypersurfaces orthogonal to the direction where the potential discontinuity may in principle lie. Although the relevant split 
to address the junction problem will be of the form $3+1$ where the first three coordinates refers to one temporal and two spatial coordinates -- the ones describing the two-dimensional hypersurfaces --  and the last one refers to the orthogonal coordinate with respect to the boundary. 
%
%
This split can be understood as totally analogous to ADM decomposition \cite{gravitation} up to signs issues, so we decided to  illustrate the ADM decomposition in the frame of $f(T)$ theories, a tool which will enable us to advance in Section \ref{sec4}. In fact, at the beginning of Section \ref{sec4} we provide the required change to go from one split (ADM) to the other $(3+1)$. Once these changes are introduced, 
both the discontinuities and Dirac-like distributions appearing in the field equations can be more easily analysed using Gaussian coordinates.

Let us then consider the usual ADM formalism and foliate the spacetime into hypersurfaces of constant time, such that the metric can be written as follows
\be 
{\rm d}s^2=N^2{\rm d}t^2-h_{ij}\left({\rm d}x^i+N^i{\rm d}t\right)\left({\rm d}x^j+N^j{\rm d}t\right)\;.
\label{ADMmetric}
\ee
Then the Ricci scalar can be rewritten in terms of the above metric:
\be
R=\Theta^2-\Theta_{ij}\Theta^{ij}+R^{(3)}-\frac{2}{\sqrt{h}N}\partial_{0} (\sqrt{h}\Theta)+\frac{2}{N}\left(\Theta N^{i}-h^{ij}N_{,j}\right)_{,i}\ ,
\label{RicciADM}
\ee
where $\Theta_{ij}$ is the extrinsic curvature defined by
\be
\Theta_{ij}=n_{\mu;\nu}\frac{\partial x^{\mu}}{dx^{i}}\frac{\partial x^{\nu}}{dx^{j}}\ ,
\label{excurvature1}
\ee
and $n_{\mu}$ is the normal vector to the hypersurface. Then, the extrinsic curvature on a hypersurface of constant time $t$ yields
\be
\Theta_{ij}=-\frac{1}{2N}\left(\dot{h}_{ij}-D_{i}N_{j}-D_{j}N_{i}\right)\ .
\label{excurvature}
\ee
Here both dot and $\partial_0$ denote derivative with respect to time $t$. In the case of $f(T)$ theories, the set of tetrads which  gives  the metric (\ref{ADMmetric}), is not unique and each choice may lead to different solutions. Consequently, for simplicity the following set of tetrads will be assumed \cite{Wu:2011kh,Blagojevic}:
\be
e^{0}_{\;\;\mu}=\left(N,{\bf 0}\right)\ , \quad e^{a}_{\;\;\mu}=\left(N^a,h^{a}_{\;\;i}\right)\ ,
\label{TetradsADM}
\ee
where $N^a=h^{a}_{\;\;i}N^{i}$ and $h^{a}_{\;\;i}$ is the set of spatial components of the tetrads. To simplify the notation, let us use $i=1,2,3$, which refer to the spatial indexes of the spacetime of the hypersurface of constant time, while we denote now $a=1,2,3$ for the tangential space, separating the 0-component. Then, the above metric is related to (\ref{TetradsADM}) through (\ref{1.1}), leading to
\be
h_{ij}=\eta_{ab}h^{a}_{\;\;i}h^{b}_{\;\;j}\ ,
\label{ADM1}
\ee
where again $a, b=1,2,3$ are raised and lowered with $\eta_{ab}$. Then, the torsion components are given by,
\bea
T^{0}_{\;\;j0}=\frac{\partial_{j} N}{N}\ , \quad  T^{i}_{\;\;j0}=\stackrel{\bullet}{D}_{j}N^i-\frac{N^i}{N}\partial_{j}N-h_{a}^{\;\;i}\partial_{0}h^{a}_{\;\;j}\ ,  \quad T^{i}_{\;\;jk}= ^{(3)}T^{i}_{\;\;jk}\ .
\label{ADM2}
\eea
The expression of the torsion scalar can be easily obtained by using the relation (\ref{RS}) and the expression of the Ricci scalar (\ref{RicciADM}):
\be
T=\Theta_{ij}\Theta^{ij}-\Theta^2-R^{(3)}+\frac{2}{\sqrt{h}N}\partial_{0} (\sqrt{h}\Theta)-\frac{2}{N}\left(\Theta N^{i}-h^{ij}N_{,j}\right)_{,i}-2D_{\mu}T^{\nu\mu}_{\;\;\;\;\nu}\ .
\label{AMD2a}
\ee
Alternatively and in analogy with the extrinsic curvature (\ref{excurvature}), we may define the extrinsic torsion as follows,
\be
\stackrel{\bullet}{\Theta}_{ij}=-\frac{1}{2N}\left(\dot{h}_{ij}-\stackrel{\bullet}{D}_{i}N_{j}-\stackrel{\bullet}{D}_{j}N_{i}\right)\ ,
\label{ADM3}
\ee
which is related to the extrinsic curvature (\ref{excurvature}) by
\be
\stackrel{\bullet}{\Theta}_{ij}=\Theta_{ij}-\frac{N^k}{2N}\left(T_{ijk}+T_{jik}\right)\ ,
\label{ADM4}
\ee
whereas the scalar torsion can be rewritten in terms of the extrinsic torsion (\ref{ADM3}):
\begin{eqnarray}
T\,&=&\,\stackrel{\bullet}{\Theta}_{ij}\stackrel{\bullet}{\Theta}^{ij} -\left(\stackrel{\bullet}{\Theta}+\frac{N^k}{N}T^{i}_{\;ik}\right)^2+T^{(3)}+\frac{N^k N^l}{2N^2}T^{ij}_{\;\;\;l}\left(T_{ijk}+T_{jik}\right)+\frac{2N^k}{N}T^{ij}_{\;\;\;k}\stackrel{\bullet}{\Theta}_{ij}\nonumber\\
&+&\frac{2}{N}\left[\frac{2}{\sqrt{-h}}\partial_{0}\left(\sqrt{-h}\frac{N^i}{N}T^{k}_{\;ik}\right)+D_{j}\left(\frac{N^iN^j}{N^2}\partial_iN\right)-T^{ij}_{\;\;\;i}\partial_jN\right]\ .
\label{ADM4}
\end{eqnarray}
In the next section, we use this formalism for foliating the spacetime into hypersurfaces which act as matching boundaries between different regions of the spacetime. Thus the ADM formalism has allowed us to express the geometrical sector of the $f(T)$ field equations in terms of geometrical quantities defined on the boundary hypersurface.
 
\section{Junction conditions}
\label{sec4}
Let us now examine the junction conditions for a general $f(T)$ action. Here we consider the case where only jumps in the matter sector are allowed, so that there are no shells/branes located at the boundary, only discontinuities in the energy-momentum tensor. In other words, the matter side of the equations does not contain any delta function in the distribution, which forbids any delta function within the gravitational sector of the field equations. In order to obtain the junction conditions, we can express the metric tensors in a Gaussian-normal frame adapted to describe the aforementioned scenario with two regions: 
\be
{\rm d}s^2={\rm d}y^2+\gamma_{ij}^*{\rm d}x^i{\rm d}x^j\ .
\label{JC1}
\ee
For this particular choice of the coordinates, the boundary between the two different spacetime regions  can be located at $y=0$, where $\gamma_{ij}^{*}$ is the induced metric or first fundamental form on the matching hypersurface and each region of the spacetime induces a particular metric. In comparison with the ADM decomposition, where the spacetime is foliated by constant time hypersurfaces, here we have not assumed any particular foliation of the spacetime, keeping $y$ as a general coordinate. Then, by following the definition (\ref{excurvature1}), the extrinsic curvature is given by,
\be
\Theta_{ij}=\frac{1}{2}\partial_y\gamma_{ij}\ .
\label{JC1a}
\ee
Note that the choice of Gaussian coordinates (\ref{JC1}) excludes crossing terms between the $y$ and $x^{i}$ coordinates in the line element (\ref{JC1}), in such a way that a $3+1$ decomposition can be followed from the previous ADM decomposition by setting $N^{i}=0$ and replacing the coordinate $t$ by the unknown coordinate $y$, while the first fundamental form $\gamma_{ij}=-h_{ij}$, as the negative sign is removed in order to keep the hypersurfaces of constant $y$ as general as possible, with no particular time-like or space-like nature. 
Then, the set of tetrads describing the metric (\ref{JC1}) can be chosen as follows 
\be
e^{y}_{\;\;\mu}=\left(1,{\bf 0}\right)\ , \quad e^{a}_{\;\;\mu}=\left({\bf 0},\gamma^{a}_{\;\;i}\right)\ ,
\label{JC1a}
\ee
which yields $\gamma_{ij}=\eta_{ab}\gamma^{a}_{\;\;i}\gamma^{b}_{\;\;j}$. The non-zero components of the torsion tensor are given by
\be
T^{i}_{\;\;yj}=\gamma^{i}_{\;\;a}\partial_{y}\gamma^{a}_{\;\;j}\ , \quad T^{i}_{\;\;jk}= \leftidx{^{(3)}}T^{i}_{\;\;jk}\ 
\label{JC1b}
\ee
and the contortion tensor and the tensor $S_{\alpha}^{\;\;\mu\nu}$ are
\[
K^{yi}_{\;\;\;\;j}=\frac{1}{2}\gamma^{ik}\partial_{y}\gamma_{kj}\ , \quad K^{ij}_{\;\;\;\;y}=\frac{1}{2}\eta^{ab}\left(\gamma^{i}_{\;a}\partial_y\gamma^{j}_{\;b}-\gamma^{j}_{\;a}\partial_y\gamma^{i}_{\;b}\right)\ , \quad K^{ij}_{\;\;\;\;k}=\leftidx{^{(3)}}K^{ij}_{\;\;\;\;k} \ ,
\]
\be
S_{y}^{\;\;yj}=\frac{1}{2}T^{ij}_{\;\;\;\;i}\ , \quad S_{j}^{\;\;yi}=\frac{1}{2}\left(\frac{\gamma^{ik}\partial_y\gamma_{kj}}{2}-\delta^{i}_{j}\gamma^{k}_{\;\;a}\partial_{y}\gamma^{a}_{\;\;k}\right)\ , \quad S_{y}^{\;\;ij}=\frac{1}{2}K^{ij}_{\;\;\;\;y}\ , \quad S_{k}^{\;\;ij}= \leftidx{^{(3)}}S_{k}^{\;\;ij}\ .
\ee
Thus, by using the above relations among this decomposition and the ADM decomposition of the previous section, the torsion scalar (\ref{ADM4}) is easily obtained in terms of the extrinsic torsion,
\be
T=\stackrel{\bullet}{\Theta}_{ij}\stackrel{\bullet}{\Theta}^{ij}-\stackrel{\bullet}{\Theta}^2+T^{(3)}\ ,
\label{JC2}
\ee
where the extrinsic torsion is given by $\stackrel{\bullet}{\Theta}_{ij}=\Theta_{ij}=\frac{1}{2}\partial_y\gamma_{ij}$. Lets remember that the foliation of the spacetime is achieved by hypersurfaces of constant $y$. Then, the junction conditions at the hypersurface $y=0$ can be obtained by analysing the field equations for a general $f(T)$ action through the equations (\ref{covEq}). The $y-y$, $y-j$ and $i-j$ components of the field equations are
\bea
&f_{T}G_{yy}&+\frac{1}{2}\left(f-Tf_{T}\right)+f_{TT}T^{ik}_{\;\;\;\;i}\partial_{k}T=\kappa^2\mathcal{T}_{yy}\ , \nn
&f_{T}G_{yj}&+f_{TT}\left(\frac{\gamma^{ik}\partial_{y}\gamma_{kj}}{2}-\delta^{i}_{j}\gamma^{k}_{\;\;a}\partial_{y}\gamma^{a}_{\;\;k}\right)\partial_{i}T=\kappa^2\mathcal{T}_{yj}\ , \nn
&f_{T}G_{ij}&+\frac{1}{2}\gamma_{ij}\left(f-Tf_{T}\right)+f_{TT}\left[S_{ji}^{\;\;\;\;k}\partial_{k}T+\left(\Theta_{ij}-\gamma_{ij}\Theta\right)\partial_{y}T\right]=\kappa^2\mathcal{T}_{ij}\ ,
\label{JC2a}
\eea
where $G_{\mu\nu}=R_{\mu\nu}-\frac{1}{2}g_{\mu\nu}R$ is the Einstein tensor whose components are given by \cite{Deruelle:2007pt}
\bea
&G_{yy}&=-\frac{1}{2}\left(\Theta_{ij}\Theta^{ij}-\Theta^2+\leftidx{^{(3)}}R\right)\ , \nn
&G_{yj}&=-\nabla_{i}\left(\Theta^{i}_{\;j}-\delta^{i}_{j}\Theta\right)\ , \nn
&G_{ij}&=\partial_y\left(\Theta_{ij}-\gamma_{ij}\Theta\right)+2\Theta^{k}_{\;i}\Theta_{kj}-3\Theta\Theta_{ij}+\frac{1}{2}\gamma_{ij}\left(\Theta_{kl}\Theta^{kl}+\Theta^2\right)+\leftidx{^{(3)}}G_{ij}\ .
\label{JC2b}
\eea
Hence, the junction conditions can now be analysed on the hypersurface $y=0$. It is straightforward to see that according to the definition of the extrinsic curvature, which contains first derivatives of the induced metric across the boundary and the expression of the scalar torsion (\ref{JC2}), the first fundamental form has to be continuous in order to avoid powers of deltas in the expression of $T$. This leads to
\be
\left[\gamma_{ij}\right]^{+}_{-}=0\ ,
\label{JC2c}
\ee
 which can also be expressed in terms of the tetrads as follows,
 \be
\left[\eta_{ab}\gamma^{a}_{\;\;i}\gamma^{a}_{\;\;j}\right]^{+}_{-}=0\ .
\label{JC2e}
\ee
Nevertheless, note that the above condition does not imply $\left[\gamma^{a}_{\;\;i}\right]^{+}_{-}=0$, as the choice of tetrads is not unique. However, the second equation in (\ref{JC2a}) imposes continuity on the induced tetrads in order to avoid delta-like distributions in the equations, which leads to the second junction condition in $f(T)$ gravities (absence in Teleparallel gravity),
 \be
\left[\gamma^{a}_{\;\;i}\right]^{+}_{-}=0\ .
\label{JC2ebis}
\ee
Furthermore, we see in the field equations (\ref{JC2a}) that the $ij$-component may also contain delta-like distributions through the term,
\be
f_T\partial_y\left(\Theta_{ij}-\gamma_{ij}\Theta\right)+f_{TT}\left(\Theta_{ij}-\gamma_{ij}\Theta\right)\partial_y T=\partial_y\left[f_T\left(\Theta_{ij}-\gamma_{ij}\Theta\right)\right]\propto \left[f_T\left(\Theta_{ij}-\gamma_{ij}\Theta\right)\right]^{+}_{-}\delta(y)\ ,
\label{JC2f}
\ee
which gives
\be
\left[f_T\left(\Theta_{ij}-\gamma_{ij}\Theta\right)\right]^{+}_{-}=0\ .
\ee
Hence, taking the trace of the above expression, the third junction condition is obtained, namely
\be
\left[f_T\Theta_{ij}\right]^{+}_{-}=0\ .
\label{JC2g}
\ee
Recall that  by assuming $f_T=1$, i.e., Teleparallel gravity, the usual junction conditions of GR are recovered, where (\ref{JC2ebis}) becomes irrelevant and (\ref{JC2g}) leads to $\left[\Theta_{ij}\right]^{+}_{-}=0$. In addition, the third junction condition is model-dependent which means that in general, matchings in GR will not be able to be reconstructed in $f(T)$ gravity. Nevertheless, in order to recover the solutions of GR, we may consider a more restrictive set of conditions that can be obtained by imposing the continuity on both terms of the left-hand side (l.h.s.) of equation (\ref{JC2f}) separately. This would involve the continuity of the second fundamental form or extrinsic curvature:
\be
\left[\Theta_{ij}\right]^{+}_{-}=0\ ,
\label{JC2d}
\ee
which consequently  implies the continuity of the extrinsic torsion (\ref{ADM4}). Then,  provided that $f_{TT}\neq 0$, the second term of the l.h.s. of  (\ref{JC2f})  contains a divergence of the scalar torsion: 
\be
\partial_{y}T=\partial_{y}T^{+}\theta^{+}(y)+\partial_{y}T^{-}\theta^{-}(y)+\left[T\right]^{+}_{-}\delta(y)\;.
\label{JC3}
\ee
which leads to the condition
\be
\left[ T\right]^{+}_{-}=0\ .
\label{JC8}
\ee
It is straightforward to check that (\ref{JC2d}) and (\ref{JC8}) form a subset of the third condition (\ref{JC2g}). In conclusion, the junction conditions for a general $f(T)$ Teleparallel theory turns out to be (\ref{JC2c}), (\ref{JC2ebis}) and (\ref{JC2g}). \\

Nevertheless, note that the set of tetrads considered here (\ref{JC1a}) is not the most general one, but just a particular choice. Let us consider now a general set of tetrads that describes the metric (\ref{JC1}),
\be
g_{\mu\nu}=\eta_{ab}\tilde{e}^{a}_{\;\;\mu}\tilde{e}^{b}_{\;\;\nu}\ .
\label{JC9}
 \ee
The first fundamental form or induced metric on the hypersurface $y=0$ is defined as follows,
\be
\gamma_{ij}=g_{\mu\nu}\frac{\partial x^{\mu}}{\partial x^{i}}\frac{\partial x^{\nu}}{\partial x^{j}}=\eta_{ab}\tilde{e}^{a}_{\;\;\mu}\tilde{e}^{b}_{\;\;\nu}\frac{\partial x^{\mu}}{\partial x^{i}}\frac{\partial x^{\nu}}{\partial x^{j}}=\eta_{ab}\tilde{\gamma}^{a}_{\;\;i}\tilde{\gamma}^{b}_{\;\;j}\ ,
\label{JC9a}
\ee
where we have defined the induced tetrads as $\tilde{\gamma}^{a}_{\;\;i}=\tilde{e}^{a}_{\;\;\mu}\frac{\partial x^{\mu}}{\partial x^{i}}$. Note that for the set (\ref{JC1a}), the induced tetrads are given by $\gamma^{a}_{\;\;i}$, defined in (\ref{JC1a}). Furthermore, as both sets of tetrads lead to the same metric, (\ref{JC1a}) and (\ref{JC9}) are connected by a Lorentz transformation,
\be
\tilde{e}^{a}_{\;\;\mu}=\Lambda^{a}_{\;\;b}e^{b}_{\;\;\mu}\ ,
\label{JC10}
\ee
while the torsion tensor for the general set of tetrads (\ref{JC9}) can be written as,
\be
\tilde{T}^{\lambda}_{\;\;\mu\nu}=T^{\lambda}_{\;\;\mu\nu}+\zeta^{\lambda}_{\;\;\mu\nu}\ , \quad \text{where} \quad \zeta^{\lambda}_{\;\;\mu\nu}=\Lambda_{a}^{\;\;b}e_{b}^{\;\;\lambda}e^{c}_{\;\;\left[\nu\right.}\partial_{\left.\mu\right]}\Lambda^{a}_{\;\;c}\ .
\label{Es16a}
\ee
Thus the torsion scalar yields
\be
\tilde{T}=T+\frac{1}{2}T^{\lambda}_{\;\;\mu\nu}\zeta_{\lambda}^{\;\;\mu\nu}+T^{\lambda}_{\;\;\mu\nu}\zeta_{\;\;\;\;\lambda}^{\nu\mu}-2T^{\rho}_{\;\;\mu\rho}\zeta^{\nu\mu}_{\;\;\;\;\nu}+\zeta\ ,
\label{Es16b}
\ee
where $\zeta=\frac{1}{4}\zeta^{\lambda}_{\;\;\;\mu\nu}\zeta_{\lambda}^{\;\;\;\mu\nu}+\frac{1}{2}\zeta^{\lambda}_{\;\;\;\mu\nu}\zeta_{\;\;\;\;\;\lambda}^{\nu\mu}-\zeta^{\rho}_{\;\;\;\mu\rho}\zeta_{\;\;\;\;\;\nu}^{\nu\mu}$. Then, the crucial term $\tilde{S}_{j}^{\;\;yi}$ of the $yi$-component of the field equations yields
\be
\tilde{S}_{j}^{\;\;yi}=S_{j}^{\;\;yi}+....-\frac{\Lambda_{a}^{\;\;b}}{4}\left(e_{b}^{\;\;i}e^{c}_{\;\;j}\partial_y\Lambda^{a}_{\;\;c}+\gamma^{im}\gamma_{jk}e_{b}^{\;\;k}e^{c}_{\;\;m}\partial_y\Lambda^{a}_{\;\;c}-2\delta^{i}_{j}e_{b}^{\;\;k}e^{c}_{\;\;k}\partial_y\Lambda^{a}_{\;\;c}\right)
\label{JC11}
\ee
The component $S_{j}^{\;\;yi}$ contains terms as $\partial_y\gamma^{a}_{\;\;j}$, which does not provide any delta-like distributions because of the second junction condition (\ref{JC2ebis}), so  in order to avoid delta-like distributions the Lorentz transformation should be the same on both sides of the boundary, $\left[\Lambda^{a}_{\;\;b}\right]^{+}_{-}=0$, or in other words, the induced tetrads of the general set $\{\tilde{e}^{a}_{\;\;\mu}\}$ has to be continuous,
\be
 \left[\gamma^{a}_{\;\;i}\right]^{+}_{-}= \Lambda_{b}^{\;\;a}\left[\tilde{\gamma}^{b}_{\;\;i}\right]^{+}_{-}=0\quad \rightarrow\quad \left[\tilde{\gamma}^{a}_{\;\;i}\right]^{+}_{-}=0.
 \label{JC12}
 \ee
Following the same procedure, it is straightforward to check that the third condition (\ref{JC2g}) has to also be satisfied when assuming the general set of tetrads (\ref{JC9}). \\

Hence, the novelty of these results with respect to other extended theories of gravity not only lies in equations (\ref{JC2ebis}) and (\ref{JC2g}), but also in the possibility of recovering the GR junction conditions from (\ref{JC2d}) and (\ref{JC8}). In the following section, we illustrate the reconstruction procedure for the Einstein-Straus model in general $f(T)$ gravity.

\section{The Einstein-Straus reconstruction}
\label{sec5}
In this section let us consider the matching of two spacetime regions 
within the framework of extended Teleparallel gravity. We focus on the so-called Einstein-Straus model \cite{Einstein-Straus}. This reconstruction consists of an empty region with a point-like mass at the center, surrounded by an expanding FLRW spacetime. The aim here is to match both regions within the general framework of $f(T)$ gravity using the junction conditions found in the previous section. In the original Einstein-Straus construction, the interior solution is described by the Schwarzschild solution:
\be
{\rm d}s^2=\left(1-\frac{2M}{r}\right){\rm d}t^2-\left(1-\frac{2M}{r}\right)^{-1}{\rm d}r^2-r^2{\rm d}\Omega_{(2)}^2\ ,
\label{ES1}
\ee
where ${\rm d}\Omega_{(2)}^2={\rm d}\theta^2+\sin^2\theta {\rm d}\phi^2$. 
%
On the other hand, the exterior region consists of a FLRW spacetime, whose metric written in the usual comoving coordinates is given by:
\be
{\rm d}\tilde{s}^2={\rm d}\tilde{t}^2-a^2(\tilde{t})\left(\frac{{\rm d}\tilde{r}^2}{1-k \tilde{r}^2}+\tilde{r}^2{\rm d}\Omega_{(2)}^2\right)\ .
\label{ES3}
\ee
Here the tilde is used to refer to those quantities defined in the FLRW region of the spacetime. 

The boundary between both regions is located at $r=R$ in the Schwarzschild spacetime and at $\tilde{r}=\tilde{R}$ in the FLRW spacetime, so that we can proceed to apply the junction conditions shown in the previous section. To do so, let's obtain the induced metric on the hypersurface $r=R$ from the Schwarzschild side, which yields
\be
\gamma_{ij}{\rm d}x^{i}{\rm d}x^{j}=\left(A^2-\frac{\dot{R}}{A^2}\right){\rm d}t^2-R^2{\rm d}\Omega_{(2)}^2\ ,
\label{ES5}
\ee
where the indexes $\{i,j\}$ refer to coordinates defined on the hypersurface $r=R$ and $A=\left(1-\frac{2M}{r}\right)^{1/2}$. Moreover, the induced metric from the FLRW region gives
\be
\tilde{\gamma}_{ij}{\rm d}x^{i}{\rm d}x^{j}\,=\,{\rm d}\tilde{t}^2-a^2(\tilde{t})\tilde{R}^2{\rm d}\Omega_{(2)}^2\ .
\label{ES6}
\ee
According to the first junction condition (\ref{JC2c}),  both induced metrics have to coincide, i.e., $[\gamma_{ij}]^{+}_{-}=0$, which gives the following relations
\be
R=a(\tilde{t})\tilde{R}\ \quad \text{and} \quad {\rm d}\tilde{t}=\frac{\sqrt{A^4-\dot{R}^2}}{A}{\rm d}t\ ,
\label{ES7}
\ee
leading to $\gamma_{ij}=\tilde{\gamma}_{ij}$. Let us now apply the second junction condition (\ref{JC2ebis}). For that purpose, the following  choice of tetrads for the Schwarzschild metric (\ref{ES1}) is preliminarily assumed, 
\be
e^{a}_{\;\;\mu}=\text{diag}\left(\left(1-\frac{2M}{r}\right)^{1/2},\ \left(1-\frac{2M}{r}\right)^{-1/2},\ r,\ r\sin\theta\right)\ ,
\label{ES2}
\ee
while the non-null components of the induced tetrads lead to
\be
\gamma^{a}_{\;\;t}=A\delta^{a}_{\; t}+\frac{\dot{R}}{A}\delta^{a}_{\; r}\ ,
\label{ES2a}
\ee
According to the second junction condition, $\gamma^{a}_{\;\;t}=\tilde{\gamma}^{a}_{\;\;t}$, so the choice of tetrads in the FLRW region is determined by such condition. Let's assume the following set:
\be
\tilde{e}^{a}_{\;\;\mu}=\begin{pmatrix}
A\ & \frac{a(t) \dot{R}}{\sqrt{(1-k\tilde{r}^2)(A^4-\dot{R}^2)}}\ & 0 \ & 0 \\
\frac{\dot{R}}{A} \ &\  \frac{a(t)}{\sqrt{(1-k\tilde{r}^2)(1-\frac{\dot{R}^2}{A^4})}}\ & \ 0\  & \ 0 \\
0 \ &\  0\ & \ \tilde{r}a\  & \ 0 \\
0 \ &\  0\ & \ 0\ & \ \tilde{r}a\sin\theta
\end{pmatrix} \ ,
\label{ES2b}
\ee
where we have assumed the change of coordinates (\ref{ES7}). Then, it is straightforward to show that the induced tetrads are continuous across the boundary. 

To analyse the third junction condition, we need to obtain the normal vector of the hypersurface from both sides of the boundary. On the vacuum side we have 
\be
n_{\mu}=\frac{A}{\sqrt{A^4-\dot{R}^2}}\left(-\dot{R}\delta^{t}_{\mu}+\delta^{r}_{\mu}\right)\ ,
\label{ES10}
\ee
while on the FLRW side, we obtain
\be
\tilde{n}_{\mu}=\frac{a}{1-k\tilde{R}^2}\delta^{\tilde{r}}_{\mu}\ .
\label{ES9}
\ee

Then, the non-zero components of the extrinsic curvature (\ref{excurvature1}) on the Schwarzschild side are
\be
\Theta_{tt}=\frac{6A^3A_{,r}-2A^7A_{,r}-2A^4\ddot{R}}{2A(A^4-\dot{R}^2)^{3/2}}\ , \quad \Theta_{\theta\theta}=\sqrt{\frac{A^6R^2}{A^4-\dot{R}^2}}\ , \quad \Theta_{\phi\phi}=\sqrt{\frac{A^6R^2}{A^4-\dot{R}^2}}\sin^2\theta\ .
\label{ES9b}
\ee
In the FLRW region, we have
\be
\tilde{\Theta}_{\theta\theta}=a(\tilde{t})\tilde{R}\sqrt{1-k\tilde{R}^2}\ , \quad \tilde{\Theta}_{\phi\phi}=a(\tilde{t})\tilde{R}\sqrt{1-k\tilde{R}^2}\sin^2\theta\,.
\label{ES9c}
\ee
In addition, the scalar torsion for the Schwarzschild tetrads (\ref{ES2}) is given by
\be
T=\frac{A}{R}\left(2A'+\frac{A}{R}\right)=\frac{2}{R^2}\ ,
\label{ES14}
\ee
while the scalar torsion for the FLRW tetrads (\ref{ES2b}) yields
\bea
\tilde{T} &=&\frac{1}{R^2(A^4-\dot{R}^2)^2}\left[2 (A^4-\dot{R}^2)^2-2\frac{R^2}{a^2}(A^4-\dot{R}^2)(A^4k+3A^2\dot{a}^2-k\dot{R}^2)\right. \nn
&&\left. +8R\sqrt{1-kR^2/a^2}A^{4}\dot{R}\sqrt{1-\frac{\dot{R}}{A^4}} -4R\sqrt{1-kR^2/a^2}A^{5}\sqrt{1-\frac{\dot{R}}{A^4}}\ddot{R}     \right]\ ,
\label{ES15}
\eea
where we have used (\ref{ES7}). Let us now apply the third junction condition (\ref{JC2g}), i.e., $\left[f_T\Theta_{ij}\right]^{+}_{-}=0$, to this setup. To do so, we may assume a particular action: for simplicity let's take $f(T)=T^m$, which leads to the condition
\be
\dot{R}=A^4\left[1-\frac{A^2}{1-kR^2/a}\left(\frac{T}{\tilde{T}}\right)^{2(m-1)}\right]\ ,
\label{ES12}
\ee
where the usual Einstein-Straus model is recovered when $T=\tilde{T}$. This is clearly not the case, as can be seen by comparing (\ref{ES14}) and (\ref{ES15}). In order to recover the Einstein-Straus result, the continuity on the torsion tensor has to be guaranteed, so that the more restrictive conditions (\ref{JC2d}) and (\ref{JC8}) have to be satisfied. However, as $f(T)$ gravity is not invariant under local Lorentz transformations, different sets of tetrads may lead to different solutions (see Ref.~\cite{Li:2010cg}). In particular, it is well known that diagonal tetrads in (\ref{ES2}) do not allow the existence of Schwarzschild solution for any action different from $f(T)=T$, which obviously reduces to Teleparallel gravity or equivalently to GR. Nevertheless, by transforming the original set of tetrads (\ref{ES2}) by applying a particular Lorentz transformation, the Schwarzschild metric might turn out to be a solution of a more general $f(T)$ action \cite{good-and-bad}. This situation also appears in the case of non-flat FLRW metrics, as pointed out in Ref.~\cite{good-and-bad,Argentinos}.  This is also the case when matching different spacetime regions and one tries to recover the usual Einstein-Straus model.

 In order to keep the second junction condition (\ref{JC2ebis}) valid, the same Lorentz transformation is applied on both sides.  Let us consider the transformation:
\begin{eqnarray}
e'^{a}_{\;\;\mu}\,=\,\Lambda^{a}_{b} e^{b}_{\;\;\mu}.
\label{ES16}
\end{eqnarray}
 As shown in the previous section, the first junction condition remains unaffected by the local Lorentz transformation due to the fact that it can be expressed in terms of Lorentz invariant quantities. Also the second junction condition (\ref{JC2ebis}) remains valid when applying the same transformation. Nevertheless,  the torsion scalar $T$ is not a Lorentz invariant, which may lead to (\ref{JC2d}) and (\ref{JC8}) by the appropriate transformation, recovering the usual result of GR for the Einstein-Straus model.

We consider the following rotation
\be
\Lambda^{a}_{\;\;b}=\begin{pmatrix}
1 & 0 & 0 & 0 \\
0 \ &\  \cos\phi\sin\theta\ & \ -\left(\sin\phi\cos\alpha+\cos\theta\cos\phi\sin\alpha\right)\  & \ \sin\phi\sin\alpha-\cos\phi\cos\theta\cos\alpha \\
0 \ &\  \sin\phi\sin\theta\ & \ \cos\phi\cos\alpha-\cos\theta\sin\phi\sin\alpha\  & \ -\left(\cos\phi\sin\alpha+\sin\phi\cos\theta\cos\alpha\right) \\
0 \ &\  \cos\theta\ & \ \sin\theta\sin\alpha\  & \ \sin\theta\cos\alpha 
\end{pmatrix} \ ,
\label{ES17}
\ee
where $\alpha=\alpha(t,r)$. Then, by keeping $\alpha$ arbitrary, the torsion scalar (\ref{ES14}) for the Schwarzschild region evaluated at the hypersurface $r=R$, turns out to be
\be
T'=4\frac{R-2M+A(R-M)\sin\alpha+AR(R-2M)\cos\alpha\ \alpha_r|_{r=R}}{R^2 (R-2M)}\,
\label{ES19}
\ee
By also applying the rotation matrix (\ref{ES17}) to the FLRW tetrads, 
the torsion scalar evaluated at $\tilde{r}=\tilde{R}=R/a$ becomes:
\bea
&&\tilde{T}'=\frac{-2}{R^2(A^4-\dot{R}^2)^2}\left\{3R^2\left(A^6-A^2\dot{R}^2\right)\frac{\dot{a}^2}{a^2}-2RA\dot{R}^3\left(2\frac{\dot{a}}{a}\sin\alpha+\cos\alpha\alpha_t\right) -\left(2-\frac{kR^2}{a}\right)\dot{R}^4\right.\nonumber\\
&& \left.+\,A^8\left[-2+k R^2/a^2-2\left(\sin\alpha+\frac{R\cos\alpha\alpha_r|_{r=R}}{a}\right)\sqrt{(1-kR^2/a^2)(1-\dot{R}^2/A^4)} 
\right]\right.\nn
&&\left. 
+\,2A^4\dot{R}\left[-2R\dot{A}\left(\sin\alpha+\sqrt{(1-kR^2/a^2)(1-\dot{R}^2/A^4)}\right)\right.\right. \nn
&& \left. \left.+\, \dot{R}\left(2-kR^2/a^2+\left(\sin\alpha+\frac{R}{a}\cos\alpha\alpha_r|_{r=R}\right)\sqrt{(1-kR^2/a^2)(1-\dot{R}^2/A^4)} 
\right)\right] \right.\nn
&&\left. +\,2RA^5\left[\sin\alpha\left(2\dot{R}\frac{\dot{a}}{a}+1\right)+\ddot{R}\sqrt{(1-kR^2/a^2)(1-\dot{R}^2/A^4)}+\frac{\dot{R}}{a}\cos\alpha \alpha_{t}\right] \right\}.
\label{ES22}
\eea
Then, it may be possible to get $T'=\tilde{T}'$ for an appropriate function $\alpha(t,r)$. Nevertheless, let us be reminded that the expression for $\alpha(t,r)$ can not be obtained analytically in general but would require numerical resources. In case of the existence of such function,  the GR junction conditions are recovered and in combination with (\ref{ES7}), equation (\ref{ES12}) leads in this case to the following well-known result \cite{Clifton:2012ry}:
\be
\frac{1}{a^2}\left(\frac{\partial a}{\partial\tilde{t}}\right)^2+\frac{k}{a^2}=\frac{2M}{\tilde{R}^3a^3(\tilde{t})}\ .
\label{ES13}
\ee
This condition requires that the FLRW side is filled with a pressureless fluid, whose energy coincides with the energy enclosed in the Schwarzschild region. Note however, that while keeping $T\neq\tilde{T}$ in (\ref{ES12}), the matching between a FLRW spacetime and the Schwarzschild solution will lead to different descriptions of the cosmic evolution that will depend on $f(T)$, while using the above tetrads and a particular expression for $\alpha(t,r)$, it gives rise to (\ref{ES13}) independently of $f(T)$. 

\section{Conclusions}
\label{Sec6}

In this paper, we have obtained  for the first time,  the junction conditions matching two regions of a spacetime within the framework of the so-called $f(T)$ theories of gravity, a natural generalisation of Teleparallel gravity.  One might have expected that since General Relativity and Teleparallel gravity are equivalent theories, 
the scenario of equivalence might be extrapolated to arbitrary functions of Ricci and torsion scalar, i.e., when comparing $f(R)$ and $f(T)$ solutions. However, we have shown above that $f(R)$ and $f(T)$ do not constitute equivalent theories as it is obvious from our expression (\ref{RS}). Therefore, matching results in $f(R)$ gravity theories do not provide in principle any hint about the phenomenology in $f(T)$ theories.

Furthermore, the $f(T)$ field equations are not invariant under local Lorentz transformations, which may lead to different solutions while working with different sets of tetrads, an element which is absent in other modified gravities. In this way, we have shown how the non-invariance of the equations affects the accomplishment of the junction conditions. For instance, the choice of the tetrads plays a crucial role due to the fact that the induced tetrads on the boundary have to be continuous (\ref{JC2ebis}). We have illustrated this fact for the Einstein-Straus model, where the matching between a Schwarzschild-vacuum solution and a Robertson-Walker region leads to a different reconstruction from the one obtained when this model is studied in pure Einsteinian gravity. In addition, the third junction condition  (\ref{JC2g}) depends upon the underlying gravitational theory, which leads to different configurations of the Einstein-Straus model unless one imposes continuity on the torsion scalar (\ref{JC8}), which may be achieved whenever a particular rotation is applied (on both sides of the boundary),  such that the General Relativity result can be recovered. A natural consequence is that the Oppenheimer-Snyder collapse scenario \cite{Oppenheimer-Snyder}, which represents the complementary matching of the Einstein-Straus configuration, i.e.,  collapsing star to form a black hole, is directly addressed since its existence and phenomenology is equivalent to the Einstein-Straus construction studied here.

In addition, it is interesting to remember that  a similar issue (existence of solutions depending upon the chosen tetrads) arises when dealing separately with Schwarzschild and Friedmann-Robertson-Walker metrics in the context of $f(T)$ theories. This may support the idea of the existence of good and bad tetrads, as pointed out in Ref.~\cite{good-and-bad}. In particular, not only the existence of the second junction condition (\ref{JC2ebis}), absent in Teleparallel gravity, affects directly the choice of tetrads, but also the third one (\ref{JC2g}) in case one wants to recover the GR results by imposing continuity on the torsion scalar, which is not invariant under local Lorentz transformations. 

In summary, we have obtained the general junction conditions for every $f(T)$ gravity, different from the ones of Teleparallel gravity, and explicitly given by (\ref{JC2c}), (\ref{JC2ebis}) and (\ref{JC2g}), which implies the continuity of the first fundamental form, the induced tetrads and the quantity $f_T\Theta_{ij}$ . These conditions depend upon both the $f(T)$ model and the choice of the tetrads. This demonstrates not only the necessity of further study to determine appropriate tetrad fields that describe the junction conditions properly, but also the necessity of analyzing possible new solutions that arise depending on the $f(T)$ Lagrangian. Specifically, the idea that the Israel equations in the presence of shells/branes could be of universal validity for theories invariant under diffeomorphisms, as claimed for fourth-order theories of gravity \cite{Senovilla:2013vra} constitutes a natural extension of this investigation.

\vspace{0.7cm}

{\bf Acknowledgments:} 
We would like to thank Jos\'e~M.~M. Senovilla and Ra\"ul Vera for useful comments about this project.
A.d.l.C.D. acknowledges financial support from MINECO (Spain) projects FPA2011-27853-C02-01,  FIS2011-23000 and Consolider-Ingenio MULTIDARK CSD2009-00064.
P.K.S.D. thanks the NRF for financial support. 
D. S.-G. acknowledges the support from the University of the Basque Country, Project Consolider CPAN Bo. CSD2007-00042, the NRF financial support from the University of Cape Town (South Africa) and MINECO (Spain) project FIS2010-15640.
A.d.l.C.D. thanks the hospitality of Kavli Institute for Theoretical Physics China (KITPC), Chinese Academy of Sciences, Beijing (China) for support during the early stages of preparation of this manuscript  as well as the Centre de Cosmologie, Physique des Particules et Ph\'enom\'enologie CP3, 
Universit\'e catholique de Louvain, Louvain-la-Neuve, Belgium and Theoretical Physics  Department, Basque Country University, Spain for assistance with the final steps of this manuscript. 
\medskip


\smallskip

\end{document}